\documentstyle[12pt,aaspp4]{article}
\begin{document}

\title{Constraining the History of the Sagittarius Dwarf Galaxy
Using Observations of its Tidal Debris}

\author{K.~V. Johnston\altaffilmark{1},
S. R. Majewski\altaffilmark{2,3,4}, M. H. Siegel\altaffilmark{2}, 
I.~N. Reid\altaffilmark{3,6} and
W.~E. Kunkel\altaffilmark{5}}

\altaffiltext{1}{Institute for Advanced Study, Olden Lane,
Princeton, NJ 08540}

\altaffiltext{2}{Dept. of Astronomy, University of Virginia,
Charlottesville, VA, 22903-0818 (srm4n@didjeridu.astro.virginia.edu, 
mhs4p@virginia.edu)}

\altaffiltext{3}{Visiting Research Associate, The Observatories of the Carnegie
Institution of Washington, 813 Santa Barbara Street, Pasadena, CA 91101}

\altaffiltext{4}{David and Lucile Packard Foundation Fellow, Cottrell
Scholar of the Research Corporation}

\altaffiltext{5}{Las Campanas Observatory, Carnegie Institution of
Washington, Casilla 601, La Serena, Chile
(skunk@roses.ctio.noao.edu)}

\altaffiltext{6}{California Institute of Technology, 105-24, Pasadena, 
CA 91125 
(inr@astro.Caltech.Edu)}

\begin{abstract}
We present a comparison of semi-analytic models of the phase-space
structure of tidal debris
with  measurements of average distances, velocities and surface densities
of stars associated with the Sagittarius dwarf galaxy (Sgr), compiled from
all observations reported since its discovery in 1994.

We find that several interesting features in 
the data can be explained by these models.
The properties of stars $\sim \pm 10-15^o$ away from the
center of Sgr --- in particular, the orientation of material perpendicular
to Sgr's orbit (Alard 1996, Mateo et al 1996,
Fahlman et al 1996, Alcock et al 1997)
and the kink in the velocity gradient (Ibata et al 1997) ---
are consistent with those expected for unbound material stripped during
the most recent pericentric passage $\sim 50$Myrs ago.
The break in the slope of the surface density seen by
Mateo, Olszewski \& Morrison (1998) at $b\sim -35^o$ can be understood
as marking the end of this material.
However, the detections beyond this point are unlikely to 
represent debris in a trailing streamer, torn from Sgr
during the immediately preceding passage $\sim 0.7$Gyrs ago,
as the surface-density of this streamer
would be too low compared to observations in these regions.
The low $b$ detections are more plausibly explained by
a leading streamer of
material that was lost more that 1 Gyr ago and has
wrapped all the way around the Galaxy to intercept the line of sight.
The distance and velocity measurements at $b=-40^o$ reported in 
Majewski et al (1999) also support this hypothesis.

We determine debris models with these properties on orbits that are consistent
with the currently known positions and velocities of Sgr
in Galactic potentials
with halo components that have circular velocities $v_{\rm circ}=140-200$ km/s. 
In all cases, the orbits 
oscillate between $\sim$12 kpc and $\sim$40 kpc from the Galactic center
with radial time periods of $0.55-0.75$Gyrs.
The best match to the data is obtained in models where
Sgr currently has a mass of $\sim 10^9 M_{\odot}$ and has orbited the Galaxy for 
at least
the last 1 Gyr, during which time it has reduced its mass by a factor of 2-3, or 
luminosity
by an amount equivalent to $\sim 10\%$ of the total luminosity of the Galactic 
halo.
These numbers suggest that Sgr is rapidly
disrupting and unlikely to survive beyond a few more pericentric passages.
These conclusions are only tentative because they rely heavily 
on the less-certain measurements of debris properties
far from the center of Sgr.  However, they demonstrate the immense potential for 
using debris
to determine Sgr's dynamical history in great detail.
\end{abstract}

\keywords{galaxies: individual (Sagittarius) -- galaxies: dwarf -- galaxies: interactions --
galaxies: kinematics and dynamics -- Galaxy: kinematics and dynamics -- Galaxy: evolution 
-- methods: analytical}

\section{Introduction}

Since its discovery 
in 1994 (Ibata, Gilmore \& Irwin 1994, 1995 ---
hereafter \cite{i94} and \cite{i95}),
the history of the Sagittarius dwarf galaxy (Sgr)
has been a matter of some debate.
Sgr is centered on the globular cluster M54 at $(l,b)=(6^o,-14^o)$
on the opposite side of the
bulge from us at 16 kpc from the Galactic center and is apparently moving 
towards the Galactic plane (\cite{i94,i95}, Ibata et al 1997
--- hereafter \cite{i97}).
The isopleths revealed by careful star count analyses
(\cite{i94,i95,i97}) are highly elongated along the direction of 
its motion, and populations plausibly associated with it have been found
even further upstream and downstream (see Alard 1996, Mateo et al 1996,
Fahlman et al 1996, Alcock et al 1997, 
Mateo, Olszewski \& Morrison 1998 and Majewski et al 1999 ---
hereafter \cite{a96,m96,f96,a97,m98} and \cite{paper1} respectively).
In addition, \cite{it99} have found an excess in the 
surface density of carbon stars aligned with the direction of elongation of
Sgr that completely encircles the Galaxy.
This suggests that Sgr is being strongly influenced by the Milky Way's
tidal field, possibly completely destroyed, yet its proximity to the Galactic 
center also indicates 
that it may have had time to circle the Galaxy many times.
If we are currently witnessing just one in a long series of encounters
it seems unlikely that it should also be the last.

\cite{jsh95} tackled this problem using $N$-body simulations
to investigate the evolution of stellar systems along highly eccentric orbits 
that would suffer few encounters during the lifetime of the Galaxy.
They found that they could fit nearly all the observations available at that
time with a wide variety of models, without requiring that
this be either the first or last passage of Sgr.
However, they did not attempt to fit the lack of velocity gradient
reported in the center of Sgr (\cite{i95}).
\cite{vw95} demonstrated that this datum alone suggests that Sgr
cannot be on a highly eccentric orbit, and their estimate for
the proper motion of Sgr has since been confirmed by observations 
(\cite{i97}), and rules out the specific models investigated by
\cite{jsh95}.
\cite{i97} argue that if Sgr is embedded in a dark matter halo,
it could in fact be impervious
to tidal destruction,
but concede that it could be being stripped and limited
by the Galactic tidal field.
This idea is illustrated with numerical simulations by
\cite{il98}.
Finally, \cite{z98} suggests that the proper motion measurements 
of Sgr and the Large Magellanic Cloud (LMC) 
are consistent with a scenario in which
Sgr was initially on an orbit much farther from the Galactic center -- 
and hence immune to the affect of the Galactic tidal field -- 
and has only recently been flung into it's current orbit through
a collision with the LMC.

If we could measure how fast Sgr is currently losing mass,
and constrain how much mass it has lost in the past, we could
make some progress towards resolving this debate about its
history and current state.
In this paper we address these questions by building 
semi-analytic models of the structure of tidal debris
that are consistent with all the current position- and velocity-space
observations
of material believed to be associated with Sgr.
In \S2.1 we describe sources and characteristics of the
data used to constrain the models
and in \S2.2 we outline the method for building the semi-analytic 
models.
We present models which agree with all the observations in
\S3 and discuss the extent of the limitations that the data
places on Sgr's orbit, current mass and mass loss history, and the
potential of the Milky Way.
We summarize our results in \S4.

\section{Tools}

\subsection{Summary of Data Sets}

In this study we have combined data from eight different
investigations that are currently available in the literature.
We take the distance 
(24 kpc), line-of-sight velocity (170 km/s) and proper motion from
\cite{i94,i95} and \cite{i97}, and the velocity trends along the
major axis of Sgr from Table 3 of \cite{i97}.
At $b=-7^o$ and $b=-4^o$
($+7^o$ and $+10^o$ away from the center of Sgr towards the Galactic disk) we
have two more distance estimates from the discovery of associated
RR Lyraes (\cite{a96,a97}), and at $b=-23^o$ we have distances both
from RR Lyraes (\cite{m96}) and main-sequence fitting
(\cite{f96}) to Sgr populations found serendipitously in studies
of the globular cluster M55.
We also have one distance and a tentative velocity measurement 
at $b=-40^o$ (see \cite{paper1} for discussion).
Finally, we use the surface density profile reported in \cite{m98},
which was produced from star counts at the position of 
the Sgr upper-main sequence in a set of 24 color-magnitude
diagrams that cover the expected locus of Sgr's debris trail
sky from $b=-24^o$ to $b=-48^o$. 

\begin{figure}
\begin{center}
\epsscale{1.0}
\plotone{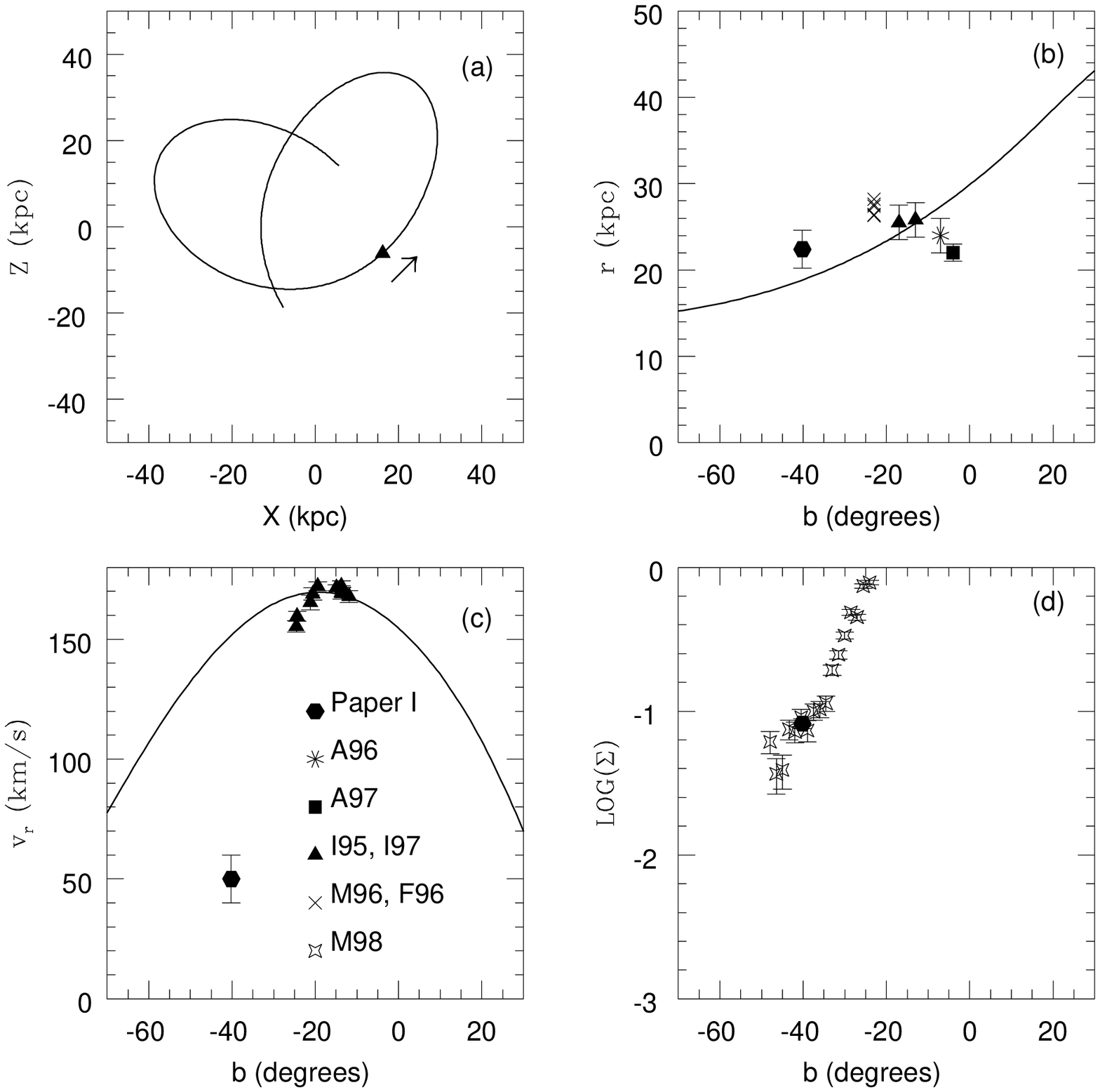}
\caption{Summary of data used in this investigation. 
Panel (a) shows the current 
position (triangle) and a possible orbit (solid line)
of Sgr projected onto the plane
perpendicular to the Galactic disk and containing the Sun at $(X,Z)=
(-8,0)$kpc. 
The arrow indicates its direction of motion.
Panels (b), (c) and (d) show the distance from the Sun, line-of-sight
velocity in a Galactic rest frame and surface density of the data 
points plotted against their Galactic latitude.
The orbit is again overlaid in Panels (b) and (c).
The sources of the data points are coded with the same
symbols in each panel, as labeled in Panel (c).}
\label{data.fig}
\end{center}
\end{figure}

The symbols in Figure \ref{data.fig} summarize this combined data set, with
the solid line in Panels (a), (b) and (c)
showing the locus of an orbit calculated from the position and velocity
of Sgr and the meaning of the symbols in each panel summarized
in Panel (c).
Panel (a) shows the position of Sgr projected onto the plane 
containing the Sun (at $(X,Z)=(-8,0)$kpc in this coordinate system)
and perpendicular to the Galactic disk.
Panels (b), (c) and (d) show the distance from the Sun, line-of-sight velocity 
in a Galactic frame of rest and (arbitrarily normalized) surface density
measurements plotted against Galactic latitude.
(Note: the surface density of stars
identified as possibly belonging to Sgr in \cite{paper1} agrees with
\cite{m98}'s estimate and this point is also included in
Panel (d) for comparison.)

When we allowed Sgr's proper motion to vary within the 
quoted error bars (\cite{i97})
and compared the different data sets both with each other and
with the loci of the orbits
we found a number of interesting features, already noted by previous authors:
the distances in the central regions of Sgr follow a line perpendicular
to the orbit rather than aligned with it;
there is a kink in the velocities measured by \cite{i97} that is hard
to reconcile with the orbit calculations alone;
although the distance measurement at $b=-40^o$
from Paper I typically lies close to the orbit,
the velocity measurement does not; and there is a break in the
slope of the Sgr star counts at $b=-35^o$.
In \S 3 we will demonstrate how all these features could 
be a natural consequence
of debris dispersal from the disruption of Sgr.

\subsection{Summary of Method}

In this section we briefly outline the 
semi-analytic method we employ to model debris dispersal ---
the details are reported in \cite{j98}.
The method predicts the full phase-space positions
and densities of mass lost at a given time from a satellite
of known mass, $m_{\rm sat}$,
and motion  orbiting in a specified parent galaxy potential, $\Phi$,
at any later time.

When a satellite disrupts, escaping stars that lose (gain) energy
move ahead of (behind) the satellite to form leading (trailing) streams of 
debris along its orbit.
In a given parent galaxy potential $\Phi$, the gain or loss of energy 
$\Delta E$ of a star can be related to its position at any time 
following the mass loss event by comparing the time period
of a circular orbit of energy $E+\Delta E$ with that of the unperturbed
satellite orbit of energy $E$ (ignoring the weak dependence
of orbital periods on angular momentum), and adjusting the azimuthal position 
of debris along the orbit accordingly.  The star's $\Delta E$ also gives an 
estimate of
its offset in distance from the orbit: $\Delta R \sim \Delta E/(d\Phi/dR)$. 
The semi-analytic method combines 
this approximate modeling of particle dynamics 
with (1) an analytic description  
of the energy scale $\epsilon=R(m_{\rm sat}/M_{Gal})^{1/3}d\Phi/dR$ 
(where $M_{Gal}$ is the mass of the parent galaxy enclosed within
radius $R$) in tidal debris; (2)
the intrinsic energy distribution (in scaled energy units $\Delta E/\epsilon$)
observed in $N$-body simulations of satellite disruption.
The method is further simplified by the observation that in
simulations of satellite destruction most mass loss occurs near the pericenter
of the satellite's orbit and hence its mass loss history 
can be approximated as a series of discrete mass loss events.
Despite these simplifications, the semi-analytic method successfully 
reproduces the density, position and velocity of streamers 
in a wide variety of 
simulations (for a given satellite
mass, orbit and approximate mass-loss history) at a tiny fraction
of the simulation's computational cost (Johnston 1998a, 1998b).
This makes the method an ideal tool for interpreting observations 
such as those described above because it allows a
quick and thorough exploration of parameter space.

Several limitations should be noted.
First, the method does not take into account the effect of dynamical
friction on the orbit of the satellite.  From \cite{bt87} 
equation 7-27, the dynamical friction timescale for 
a $10^9 M_{\odot}$ satellite orbiting at $R=$15 kpc from the center of the
Galaxy is $\sim 5$ Gyrs.  Since we consider $\le 1.5$ Gyrs in
the evolution of satellites of mass $\le 10^9 M_{\odot}$
on orbits where most time is spent at $R>15$ kpc, we expect this
simplification to be unimportant.  Moreover,
our interpretation of the data is only intended to be approximate.
However, once our knowledge of the structure and extent of
debris associated with Sgr becomes detailed enough to warrant a more
specific interpretation of its history, the effect of dynamical friction should
be included.  

Second, the method is most accurate for small satellites,
with masses such that $(m_{\rm sat}/M_{Gal})^{1/3} \ll 1$
--- for our own Galaxy this suggests it should only be used to study
satellites, such as Sgr, with $m_{\rm sat} \le 10^9 M_\odot$.
Finally, the method was developed and tested under circumstances 
where the satellite's orbit
was eccentric but not radial, and where the satellite's structure was 
supported by random motions.  Sgr satisfies both these conditions.

\section{Results}

\subsection{General Approach}

\begin{figure}
\begin{center}
\epsscale{0.5}
\plotone{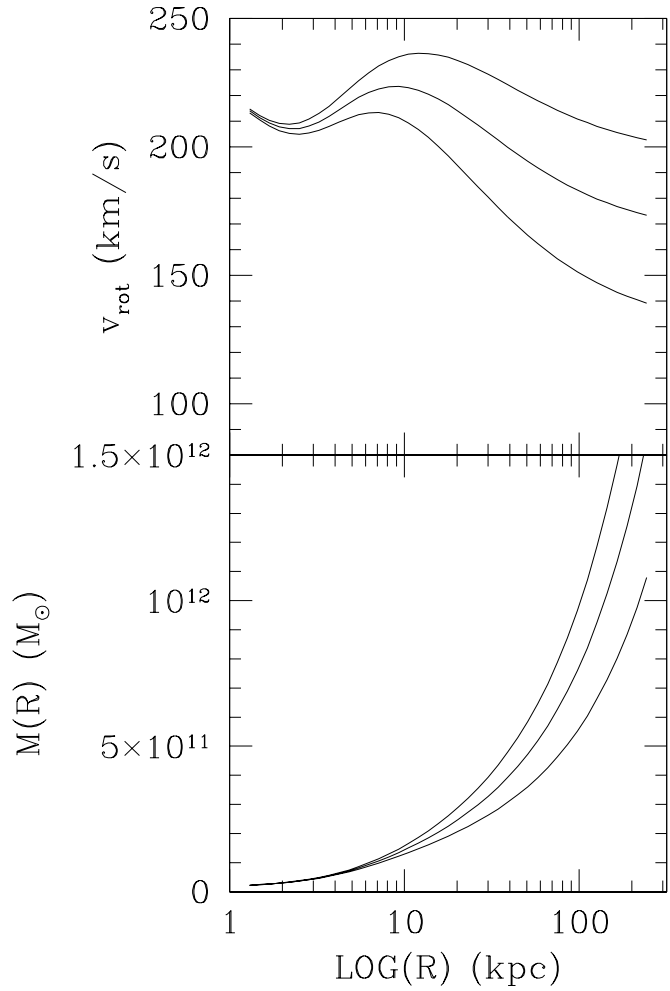}
\caption{Top panel --- the velocity of a circular orbit in the
disk plane at distance $R$ from the Galactic center
for the Galactic model given in equation (\ref{mw})
with $v_{\rm circ}=140,174,200$ km/s.
Bottom panel --- the mass enclosed within distance $R$ for the same models.}
\label{mw.fig}
\end{center}
\end{figure}

Our aim is to explore whether the combined data sets are consistent with the 
semi-analytic models 
of debris dispersal from satellite disruption and how much such a model can 
tell 
us about 
Sgr's history, orbit and the Milky Way's potential.  We do not attempt to 
use an 
automated routine
to find the ``best fit'' model since we believe that neither the extent of the 
data nor the accuracy of
the models at present justify this level of sophistication.
We instead simply compare the data and the models by visual inspection, 
varying 
four parameters:
\begin{enumerate}
	\item $f$ --- the fraction of mass lost on each pericentric passage.
We assume that mass loss occurs instantaneously at pericenter and that
$f$ is constant (as is approximately true in simulations).
	\item $m_{\rm Sgr}$ ---
the mass of material still bound to Sgr and
centered around M54.
	\item $v_{\rm t}$ --- Sgr's velocity transverse to the line of sight.
We take Sgr's orbit to be exactly polar, fix its line-of-sight velocity 
to its observed value (170 km/s), and allow $v_{\rm t}$ to vary within the
error-bars on the proper motion measurement quoted in I97.
In fact, Sgr's motion parallel to the Galactic plane is poorly 
constrained by current observations (\cite{i97}), although the surface density
contours do indicate that it must be on a near-polar orbit, and
hence it is likely to be small.
We will discuss the effect of non-polar orbits in flattened halo potentials
in \S 3.5.
	\item $v_{\rm circ}$ --- the circular velocity of the dark matter halo,
as defined in equation (\ref{halo}) below.
We integrate Sgr's orbit in Milky Way potentials of the
form $\Phi=\Phi_{\rm bulge}+\Phi_{\rm disk}+\Phi_{\rm halo}$
where
\begin{eqnarray}  \label{mw}
\Phi _{{\rm bulge}}&=&-{{GM_{{\rm bulge}}}\over {\sqrt{R^2+z^2}+c}},\cr 
\Phi _{{\rm disk}}&=&-{{GM_{{\rm disk}}}\over {\sqrt{R^2+(a+\sqrt{z^2+b^2})^2}}}
,\cr 
\label{halo}
\Phi _{{\rm halo}}&=&{\frac{v_{{\rm circ}}^2}2}\ln (R^2+z^2/q^2+d^2).
\end{eqnarray}
Here $R$ and $z$ are cylindrical coordinates aligned with the Galactic
disk and  
the numerical values adopted for the parameters are 
$M_{{\rm disk}}=1.0\times 10^{11},
M_{{\rm bulge} }=3.4\times 10^{10},
a=6.5,b=0.26,c=0.7$ and $d=12.0$, where masses are
in $M_{\odot }$, velocities are in km/s and lengths are in kpc.
We vary $v_{\rm circ}$ between 140 km/s and
200 km/s and set $q=1$ to consider spherical halo potentials.
The effect of flattened halo potentials is discussed in
\S 3.5.
The top panel of Figure \ref{mw.fig} plots the rotation curves
for the combined potential $\Phi$ with
$v_{\rm circ}=140, 170$ and 200 km/s, and the bottom panel shows
the mass enclosed for these choices within a given radius.
\end{enumerate}

\begin{figure}
\begin{center}
\epsscale{1.0}
\plotone{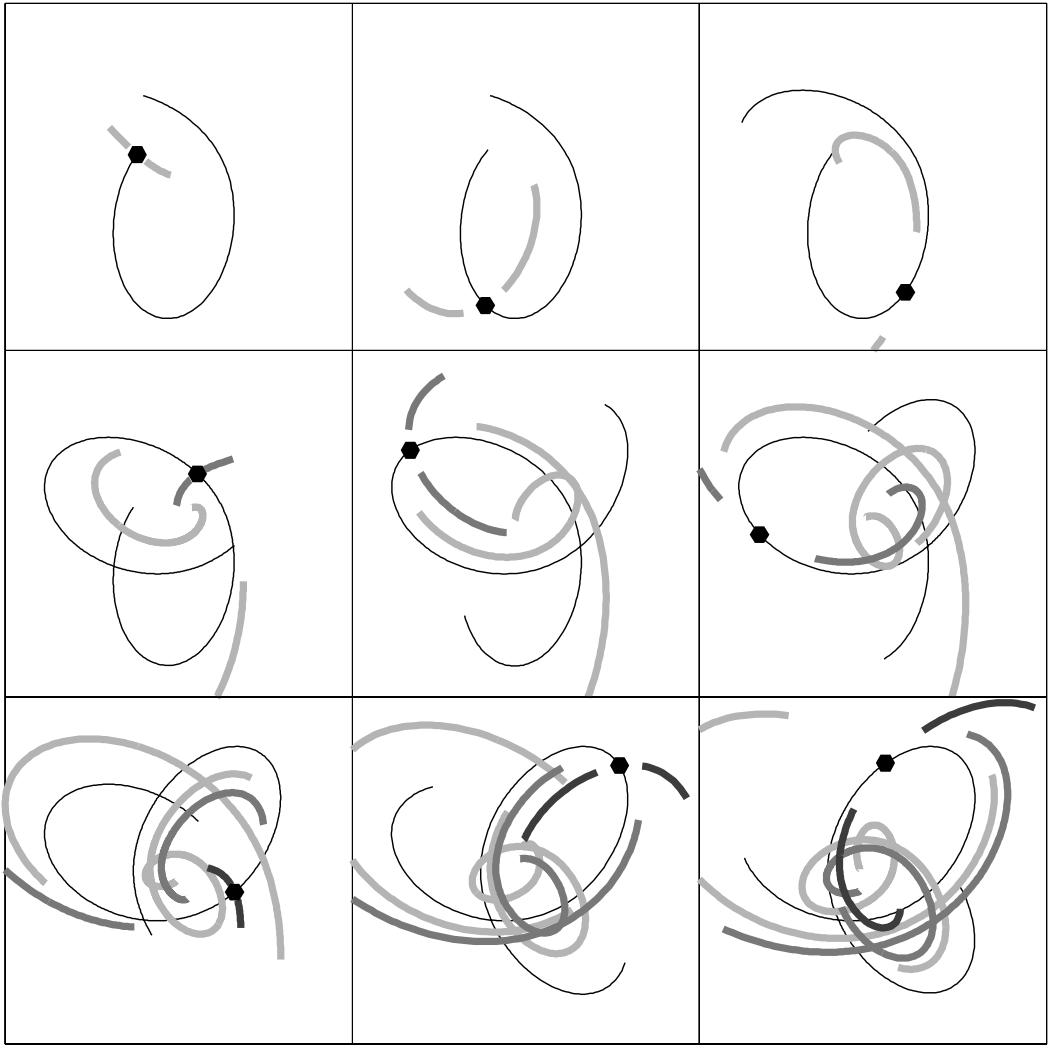}
\caption{Evolution of debris lost from a satellite over three pericentric
passages projected onto the plane
perpendicular to the Galactic disk and containing the Sun.
In each panel the solid line shows the past and future path of the satellite,
and the hexagon shows its position. The panels are equally spaced
by 0.22 Gyrs in time, centered on the Galaxy and 100kpc on a side.
The grey lines show the positions of the streamers.
The bottom left hand panel corresponds to the position and velocity of Sgr 
today.}
\label{evol.fig}
\end{center}
\end{figure}

In Figure \ref{evol.fig} we illustrate the general characteristics of 
debris dispersal by following a semi-analytic model for
the evolution of a satellite with parameters
$(f,m_{\rm Sgr},v_{\rm t},v_{\rm circ})=(0.3,10^9M_\odot,242 {\rm km/s},
174 {\rm km/s})$ over three
pericentric passages.
Each panel is centered on the Galaxy and 100 kpc on a side.
The solid line shows the past and future path of the satellite,
and the hexagon shows its position. 
The left hand panels correspond to instants shortly after each pericentric
passage (with the bottom left hand panel corresponding to Sgr's current
position and velocity) and the panels in each row are equally spaced
between these points in time.
The grey ribbons show the location of the debris lost on each 
pericentric passage.
The darkest and shortest pair of ribbons in the final panel correspond to the 
leading and trailing streamers from the most recent passage (hereafter 
referred to as passage 0 or by $n_{\rm pass}=0$)
and the paired streamers from the preceding passages (
hereafter referred to by $n_{\rm pass}=-1,-2...$)
are progressively lighter and longer.
Note that the large $m_{\rm Sgr}$
means that the tidal scale $(m_{\rm Sgr}/M_{\rm Gal})^{1/3} \sim 1/10$
is large and so the leading (trailing) 
streamers are on orbits that 
come (go) considerably closer to (further from)
the Galactic center than the satellite itself.
Each of these streamers will have a width less than, but 
comparable to the amount by which it is offset from the orbit so there will
considerable overlap between the populations whose average positions
are indicated by the grey lines.

In Figure \ref{model.fig} we repeat Figure \ref{data.fig}, but with 
the predictions of the semi-analytic model shown in Figure \ref{evol.fig}
overlaid.
In the following subsections we use this  to illustrate
the general successes of
using the dynamics of tidal debris to explain the data sets (\S 3.2).
We then go on to discuss whether the current observations
can place any stringent limits on the varied parameters (\S 3.3),
and outline the implications for Sgr's mass
loss history (\S 3.4). 
Finally, we examine to what extent our discussion will remain
valid for non-polar orbits in flattened halo potentials
(\S 3.5).

\subsection{Successes of the Models}

\begin{figure}
\begin{center}
\epsscale{1.0}
\plotone{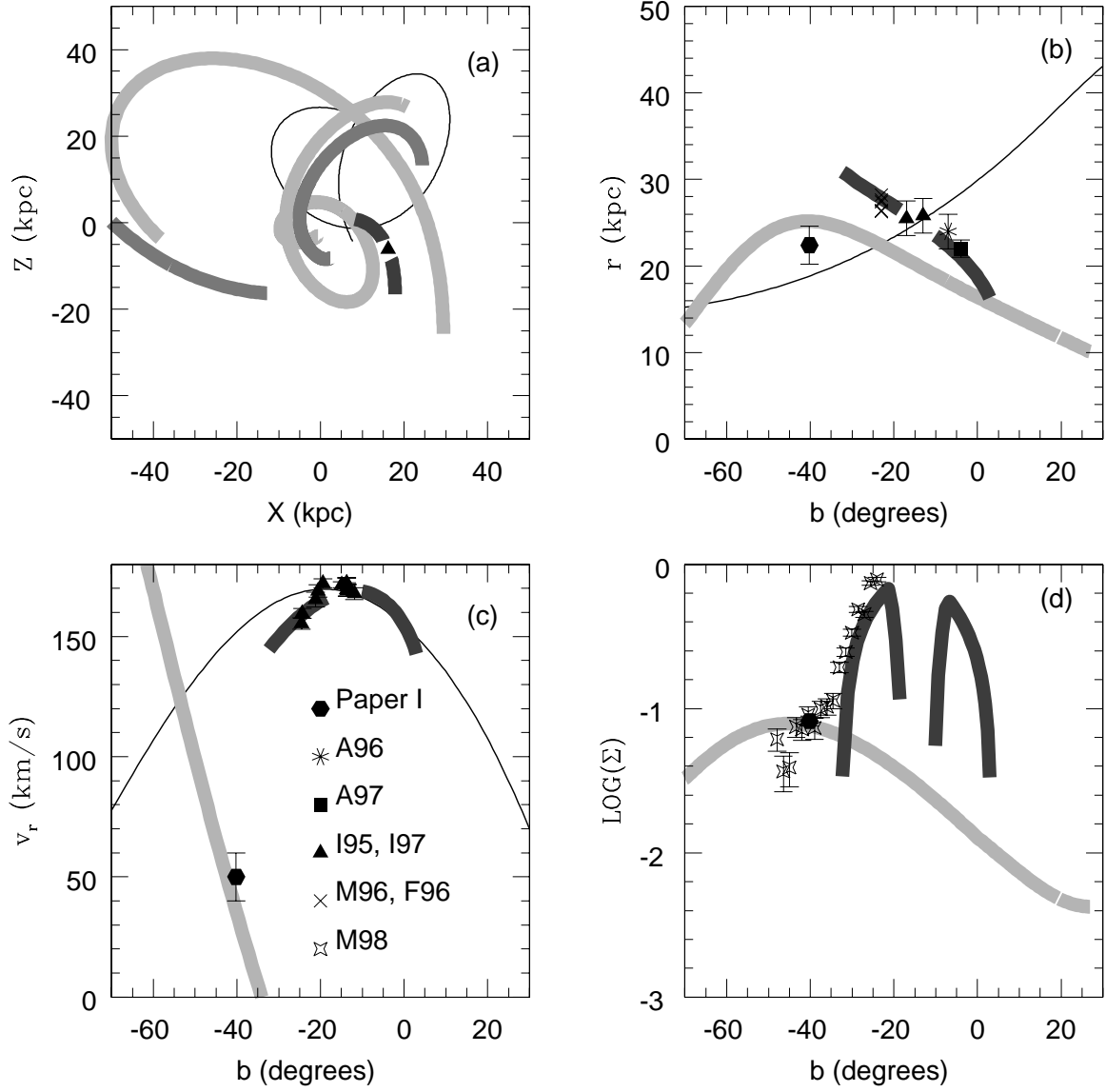}
\caption{As Figure \ref{data.fig}, but with the semi-analytic model
also overlaid. Each pair of grey lines represent the locus of 
a pair of tidal streamers, with the darkest pair being torn from the
satellite on the most recent pericentric passage, 50 Myrs ago.}
\label{model.fig}
\end{center}
\end{figure}

The location of the streamers (grey ribbons)
and the orbit (solid line) projected onto the
plane containing the Sun and perpendicular to the Galactic disk
are shown in panel (a) of Figure \ref{model.fig}.
Note that the streamers from the most recent pericentric passage
($n_{\rm pass}=0$)
are nearly perpendicular to the orbital path, and that those from the 
preceding passages ($n_{\rm pass}=-1,-2...$)
are clearly separated by a noticeable gap
ahead and behind the satellite along its path.
This general configuration is apparent in all models, and results from
Sgr being only slightly past pericenter in orbital phase ---
once it reaches apocenter the streamer from passage 0 will spread far enough 
to overlap in phase with the older ones and the debris will
be continuous (see Figure \ref{evol.fig}). 
However, currently this means that as we trace the debris
back along the orbit of Sgr beyond the extent of material unbound 
during passage 0
we will not immediately detect material
from passage $-1$, but rather
debris that has wrapped entirely around the Galaxy and hence must have been 
torn from Sgr on passages with $n_{\rm pass}\le -2$,
as illustrated by the position of the lightest leading 
streamer in the Figure.

In panel (b) of Figure \ref{model.fig}
we plot the distances to the streamers as
seen from the Sun in the direction of the center of the Galaxy as a function
of Galactic latitude.
Note that the streamers from passage 0 
agree very well with the peculiar orientation of material
reported in \cite{a96,m96,f96} and \cite{a97}.
The distance estimate we found for our field
further down the streamer in Paper I agrees with the prediction for material
in the leading streamer from passage $-2$ that has wrapped
entirely around the Galaxy to intersect our line of
sight.  This point serves as a discriminant between the streamers:
trailing streamers are offset from the satellite's path
to greater distances and should not be
observed so close to the Galactic center; and
material in leading streamers from passages with $n_{\rm pass}\le -3$
will lie progressively
closer to the satellite's orbit (and further from us
along the line of sight close to Sgr)
at a given orbital phase.
These arguments apply to all models.

In panel (c) of Figure \ref{model.fig} 
we plot the line-of-sight velocity as seen from
the Sun in the Galactic rest-frame.
The model agrees
with the kink in the velocity gradient for the points
close to the center of Sgr that was  
commented on by \cite{i97}.
The velocity reported at $b=-40^o$ in Paper I
is too low to agree with a direct extrapolation of the 
orbit of Sgr over 20$^o$ within the stated error-bars of the proper motion
in any of our Galactic models. 
However, it can be explained consistently within the model where
this material is part of a leading streamer from passage $-2$.
For a given Galactic model, the distance and velocity of this
point, provides the strongest constraint on
the orbit of Sgr.

In panel (d) of Figure \ref{model.fig} we show the surface density of
each streamer, normalized so that 
the streamer from passage 0 has the
same amplitude as the data in the region reported in M98.
In all models, there is a gap of $10^o-15^o$
between the leading and trailing streamers
from passage 0 corresponding to the location of
mass still bound to Sgr.
The models suggest that the break in slope of the
surface density at $b\approx -35^o$ corresponds to the
end of the streamer from passage 0.
The location of this break is dependent on $m_{\rm Sgr}$
and the density of debris beyond this point is
sensitive to $f$.

\subsection{Limits on Parameters}

In order to investigate to what extent the data could limit the parameters
we looked at a set of potentials with halo components having
$140{\rm km/s} \le v_{\rm circ} \le 200$ km/s.
In each potential we found the value of $v_{\rm t}$
that allowed the closest fit (by visual inspection) 
to all the distance and velocity data, except for the
tentative velocity measurement at $b= -40^o$ (see \cite{paper1} for discussion
of this point).
We considered debris from the last 3 passages alone
and did not look at debris from earlier passages because 
(as noted in the previous section) these streamers 
would be too distant to fit
the point measured at $b=-40^o$ in \cite{paper1}.
In all potentials we could find a value of $v_{\rm t}$
that provided a reasonable fit to these data within the error bars of the
proper motion measurement reported in I97.
As $v_{\rm circ}$ was increased from 140 km/s to 200 km/s, the necessary
$v_{\rm t}$ increased from 205-275 km/s (with a range of about
$\pm 10$ km/s about each value) and the time periods
and time interval since the most recent pericentric passage decreased
from 0.75-0.55 Gyrs
and 60-40 Myrs respectively.
However, the pericenters and apocenters of the orbits
orbits lay in narrow ranges (12-14 kpc and 40-43 kpc).
We conclude that although the current observations 
do not significantly refine
our model of the Galactic potential, the proper motion
measurement of Sgr, nor the prediction for the time period of its orbit, 
they do constrain the orbital shape.
If we knew the proper motion of Sgr with greater precision we could use 
the properties of the debris to 
constrain the Galactic potential more tightly.

Note that the model that fits all the distance and \cite{i97}'s velocity
measurements also provides a good match to the uncertain velocity
measurement reported in \cite{paper1}, even though it was not
considered as a constraint. This success supports
the tentative identification in \cite{paper1}
of a group of stars moving at
this velocity as indeed being associated with Sgr.

Once we determine a value of $v_{\rm t}$ in each potential 
we can use the surface density measured in M98 to constrain $m_{\rm Sgr}$.
Increasing $m_{\rm Sgr}$ causes the length of each streamer
to increase (approximately $\propto m_{\rm Sgr}^{1/3}$).
In all potentials considered we found that $m_{\rm Sgr} \sim 10^9
M_\odot$ in order for the streamer from passage 0 to extend to the point where
the slope in the surface density profile in 
M98 changed. M98 estimate a total luminosity associated with Sgr of 
$\le 8 \times 10^7 L_\odot$ (i.e. including their detections) so this suggests
an overall mass-to-light ratio of $\ge 10$, which is consistent with
observations of other dwarf spheroidal satellites in orbit around
the Milky Way, and with the independent estimates for its mass by I97.
However, this large mass is somewhat inconsistent with
the measured velocity dispersion of $\sim 10$ km/s along the entire
length of Sgr (\cite{i97}).
This inconsistency could arise because the finite size of the satellite
is ignored in the semi-analytic models, where the debris is effectively being
released from the satellite's center.
This simplification would affect the predictions for
the length of the tidal streamers from the most recent pericentric passage 
(passage 0) whose size is comparable to 
that of the satellite itself, but
should be unimportant for the much longer 
streamers from earlier passages.

Finally, we found that $f$ 
was required to be in the range $0.2 \le f \le 0.3$
in order for the relative amplitude of debris 
at either side of the break in the slope of the surface density
profile approximately to be fit by material in the streamers from
passages 0 and $-2$. 

\subsection{Mass Loss History and Future Survival}

One question that has been addressed repeatedly since Sgr's discovery 
is whether we are witnessing its death throes or will Sgr survive another
passage.
The models that best fit the data described in the previous section
had 
$0.2 \le f \le 0.3$, which means 
that Sgr has lost between half and two-thirds of its mass
over the last three pericentric passages (or 1.1-1.5 Gyrs depending on the 
Galactic potential).
We can crudely check this picture by assuming that our success in
modeling the distance and velocity features means that Sgr is 
dominated by unbound material beyond at least $10^o$ from its center.
M98 report that the surface brightness (in mag/arcsec$^{-2}$)
of their points between $10^o$ and $20^o$ from Sgr can be fitted by a function
$\Sigma_v \propto 0.23 \theta$ (where $\theta$ is the
separation from the center of Sgr in degrees)
which agrees with the central surface brightness of Sgr
when extrapolated inwards.
If mass follows light,
this suggests that the mass surface density
will fall as $\Sigma \propto \exp(-\theta/10)$.
Hence, if Sgr and its debris have roughly constant
width along its orbit the ratio of the mass 
between $10^o\le \theta \le 20^o$ (i.e. in the streamers from
passage 0)
to that within 
$\theta=10^o$ will be $\sim exp(-1)=0.37$, in rough agreement 
with our assumed fractional mass loss rates.
Such a large mass loss rate 
rules out the picture where Sgr is merely distorted by
the Milky Way's tidal field with very little evolution
and suggests that Sgr is unlikely to survive many more pericentric 
passages.
This mass loss rate corresponds to a decrease in Sgr's luminosity in the last
1.1-1.5 Gyrs alone that is of order 10\% of the total luminosity
of the stellar halo ($\sim 10^9 L_{\odot}$, see \cite{free96}).

\subsection{Halo Flattening and Non-Polar Orbits}

We have so far limited our discussion to polar orbits in 
potentials with exactly spherical halos.
Non-polar orbits in a 
flattened halo potential (i.e. setting $q < 1$ in equation [\ref{halo}]) 
will precess and disturb the alignment of the leading and trailing streamers.
In principle, we could use the apparent precession of the debris 
at different points along the streamers to place limits on
both the direction of the proper motion and the degree of flattening of the 
Galaxy's potential. However, the strongest constraints on these
quantities would come from detections (or lack of detections) 
farthest from the center of Sgr (e.g. 
\cite{m98,paper1}) which are currently the 
most uncertain. Hence, we do not try to place limits on 
these dimensions of parameter-space, and instead simply
examine to what extent our discussion of debris evolution in
a spherical halo will remain valid in the non-spherical case.

\begin{figure}
\begin{center}
\epsscale{0.75}
\plotone{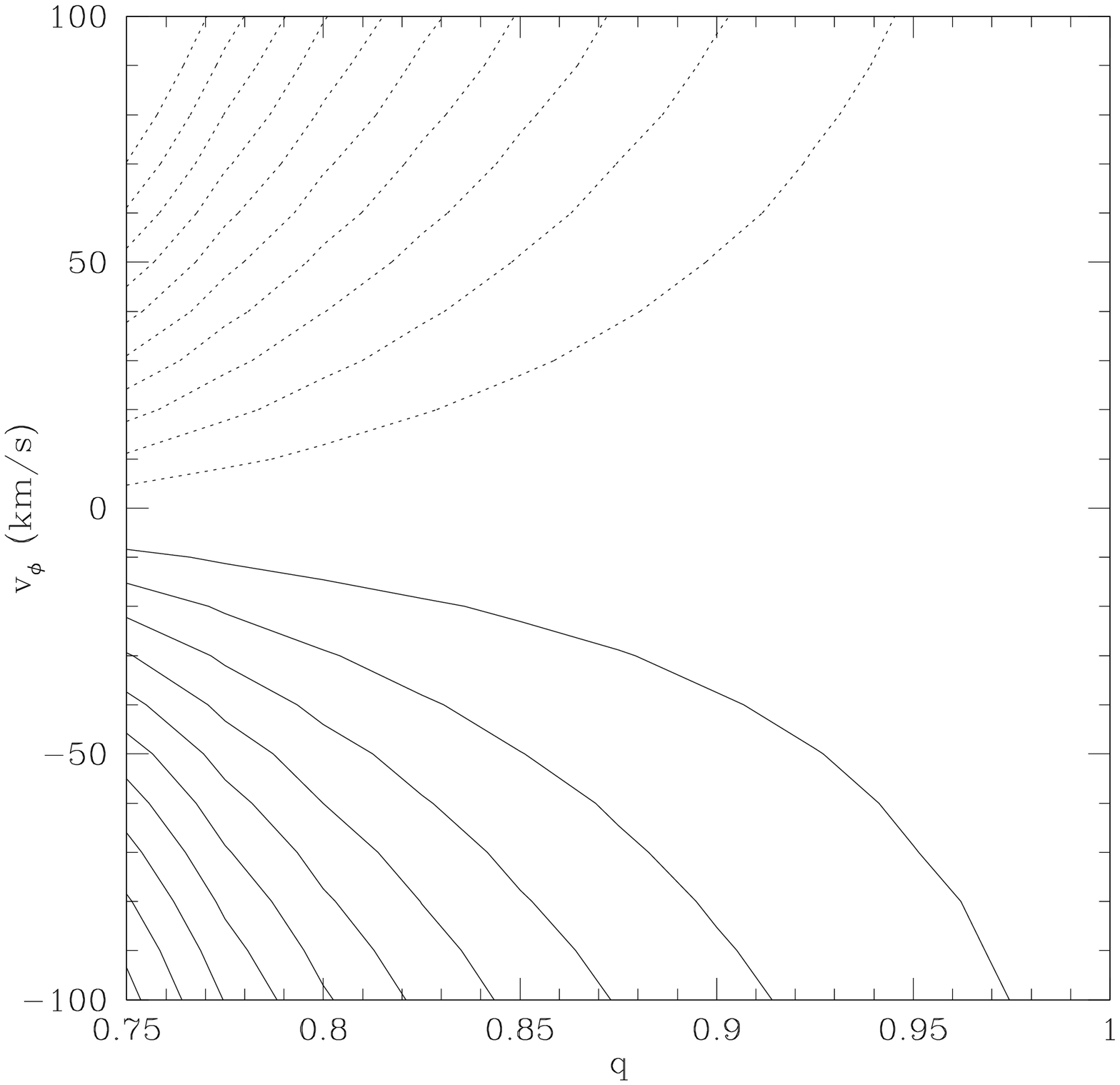}
\caption{Positive (solid) and negative (dotted) contours
of angular precession of orbit, as a function of the
flattening $q$ of the halo potential and the velocity of Sgr
perpendicular to the
line of sight and
parallel to the plane of the Galactic disk $v_\phi$.  The
levels are spaced by 5 degrees.}
\label{flat.fig}
\end{center}
\end{figure}

Figure \ref{flat.fig} contours the projected angular difference
between Sgr and a point $2\pi$ ahead along it's orbit.
The orbits were run
in halo potentials with different $q$ and with initial
radial velocity and motion perpendicular to the
Galactic disk set by \cite{i97}'s observations, but with motions
perpendicular to the line of sight and parallel to the Galactic disk
$v_\phi$ allowed to vary.
The positive (solid) and negative (dotted) contours are
spaced by 5 degrees.
This plot provides upper limits to the  apparent
orbital precession of material in the leading streamer since
debris on orbits more tightly bound to
the Galaxy than Sgr would lie closer to the Galactic center 
than it's orbit and
hence have smaller angular separations from Sgr when seen
from the Sun in projection. Figure \ref{flat.fig} clearly
cautions that our identification of \cite{m98}'s and \cite{paper1}'s
detections as material in the leading streamer could be
wrong if $q < 0.9$ and $|v_{\phi}| > 20$ km/s.
Although this condition appears to be rather restrictive, note 
that this modest potential flattening translates to
an axis ratio $\sim 0.7$ in the corresponding density distribution
(see Binney \& Tremaine 1987, equation 2-55b).
In comparison,
a recent study combining limits from stellar-kinematics and
from the flaring of gas out of the Galactic plane constrains the
dark matter halo to have an axis ratio in density of $0.75\pm 0.25$
(\cite{om98}).

\section{Summary and Conclusions}

In this paper, we have combined all observations of Sgr that
have been reported in the literature since 1994 and
directly compared them to semi-analytic models of the dispersal of
material once it becomes unbound from a satellite.
By varying the circular velocity of the halo component of our
Milky Way potential along with Sgr's 
transverse velocity, current mass and mass loss rate,
we have found that:
\begin{enumerate}
	\item
The current data cannot constrain the Milky Way potential and
Sgr's transverse velocity independently, but do favor 
specific orbital characteristics:
the orbits of the best fit models all had
pericenters $\sim 13$kpc, apocenters $\sim 41$kpc, radial time periods
$\sim 0.55-0.75$Gyrs 
and time since the most recent pericenter $\sim 40-60$ Myrs.
If the proper motion of Sgr were known with greater precision, the
data could be used to constrain the Galactic potential more tightly.
Note that \cite{md99} also found that poor knowledge of proper motion
was the limiting factor to
using globular cluster tidal tails to measure the Galactic 
potential.
	\item
All the peculiar features in the data could be explained with
the dynamics of unbound material:
	\begin{itemize}
		\item  The orientation of material perpendicular to
Sgr's orbit close to its center
reported in \cite{a96,a97} and \cite{paper1}, and the kink in the radial
velocities reported in \cite{i97} are consistent with
the predicted locus of debris from the most recent pericentric
passage.
		\item The radial velocity and distance 
measurements of material at $b=-40^o$ reported in Paper I
correspond to those expected for material in a leading streamer that
has wrapped entirely around the Galaxy to intercept our line of sight.
The density of this material in this region dominates over trailing material
because there is a clear gap between material stripped from the
most recent and immediately preceding pericentric passages.
		\item The break in the slope of the surface density profile
reported in \cite{m98} naturally occurs in our model at the transition
between material in the trailing streamer from the most recent
passage and the leading streamer that has wrapped around the Galaxy.
	\end{itemize}
	\item
Models that are consistent with the data have current masses of
Sgr of $\sim 10^9 M_{\odot}$, and mass loss rates such that Sgr
was 2-3 times larger $\le 1.5$Gyrs ago.
This mass loss rate corresponds to a decrease in Sgr's luminosity 
that is of order 10\% of the total luminosity
of the stellar halo and suggests
Sgr is unlikely to survive many more pericentric 
passages.
\end{enumerate}

We consider the general scenario outlined in our first two points
to be fairly robust, provided the halo potential is not highly
flattened (see \S 3.5).
The specific model for the mass and mass loss rate
in the third point is more tentative.
Future observations should reveal the history of Sgr in greater detail.
Velocity measurements along 
the entire length of \cite{m98}'s data set would clearly confirm or rule out
our identification of this material as part of the leading streamer
and distance estimates 
would indicate whether we are indeed seeing only 
material wrapped around the Galaxy from the pericentric passage
$\sim 1.1-1.5$ Gyrs ago (i.e. $n_{\rm pass}=-2$), or if we might also 
be seeing material (a few kpc more distant)
from  earlier passages.
The absence of debris from these earlier passages in M98's data set 
could provide support for
Zhao's (1998) model of Sgr's history, which requires Sgr to be on a
much less tightly bound orbit before
suffering a collision with the LMC 2 Gyrs ago.
This scenario limits the number of close encounters
with the Milky Way that Sgr has undergone, and hence the 
number of streamers from different passages associated with it.
On the other hand,
if Zhao's model can clearly be ruled out, we instead face the task of 
building a consistent picture in which Sgr has orbited the Galaxy
many times and is only now being completely destroyed.

Additional detections (ideally with distance, velocity and surface
density estimates) or lack of detections of stars 
associated with Sgr in fields in the Northern Galactic hemisphere
(e.g. \cite{paper1}) or radial velocity measurements
along \cite{m98}'s data set would allow us to pin down Sgr's path, which in
turn would more precisely constrain both the shape and depth of the Galactic
potential and Sgr's proper motion. 
Unfortunately, the converse of this statement is that
our poor knowledge of Sgr's motion parallel to the
Galactic disk and our uncertainty about the degree to which the
Galaxy's potential is flattened mean that any predictions of where to 
find such streamers are very imprecise.
However, the recently discovered carbon star stream (Irwin \& Totten 1999)
could provide a starting point for such studies ---
a preliminary analysis of the carbon star data set alone 
suggests that the halo potential
is near spherical (private communication from Rodrigo Ibata).
Further into the future the GAIA satellite (Global Astrometric Imager for
Astrophysics --- to be launched in 2009) will provide proper motions
of all stars brighter than 20th magnitude with $\sim 5-200 \mu as/$yr
accuracy and allow us easily to distinguish Sgr's debris 
across the sky from foreground
contaminants.

Finally, we caution that the
results reported here are only intended as a guide to the
probable history of Sgr and that
more detailed velocity and distance information along its streamers
is needed to come to firmer conclusions.  The semi-analytic models
are not exact, and, with more data to use as constraints,
any results should ideally be confirmed using N-body simulations
that also
take into account the effect of dynamical friction.
Nevertheless, this study illustrates the tremendous power
of using debris, rather than the properties of bound material,
to understand the history of satellite galaxies.

\acknowledgements
KVJ would like to thank David Spergel for helpful comments on this
manuscript, Mario Mateo for pointing out additional data
sets that should be included in the discussion and Rodrigo
Ibata for his constructive referee's report.
Her work was supported by funds from the Institute for Advanced
Study.  SRM is supported by a Cottrell Scholar Award from The
Research Corporation, and a David and Lucile Packard Foundation
Fellowship.


\clearpage

\begin{thebibliography}{}

\bibitem[A96]{a96} Alard, C. 1996, \apjlett, 458, L17 (A96)

\bibitem[A97]{a97} 
Alcock, C., Allsman, R., Alves, D. Axelrod, T., Becker, A., 
Bennett, D., Cook, K., Guern, J., {\it et al.} 
1997, \apj, 474, 217 (A97)

\bibitem[Binney \& Tremaine (1987)]{bt87} 
Binney, J. \& Tremaine, S. 1987, Galactic Dynamics
(Princeton University Press, Princeton)

\bibitem[F96]{f96} 
Fahlman, G., Mandushev, G., Richer, H., Thompson, I. \& 
Sivaramakrishnan, A. 1996, \apjlett, 459, L65 (F96)

\bibitem[Freeman 1996]{free96}
Freeman, K. C. 1996, in The Formation of the 
Galactic Halo: Inside and Out, ASP Conf. Ser. Vol 92, eds. H. Morrison \&
A. Sarajedini, (ASP, San Francisco), p. 3 

\bibitem[Ibata \& Lewis (1998)]{il98}
Ibata, R. A. \& Lewis, G. F. 1998 \apj, 500, 575

\bibitem[I94]{i94} 
Ibata, R., Gilmore, G. \& Irwin, M. 1994, Nature, 370, 194 (I94)

\bibitem[I95]{i95} 
Ibata, R. A., Gilmore, G. \& Irwin, M. J. 1995, \mnras, 277, 781 (I95)

\bibitem[I97]{i97} Ibata, R., Wyse, R., Gilmore, G., Irwin, M. \& Suntzeff, N. 
1997, \aj, 113, 634 (I97)

\bibitem[Irwin \& Totten (1999)]{it99}
Irwin, M.J. \& Totten, E. 1999, in preparation

\bibitem[Johnston (1998a, 1998b)]{j98} 
Johnston, K. V. 1998a, \apj, 495, 297

\bibitem[]{}
Johnston, K. V. 1998b, to appear in ASP Conference Ser,
``The Galactic Halo'', eds. Putman and Gibson

\bibitem[Johnston, Spergel  \& Hernquist (1995)]{jsh95} 
Johnston, K. V., Spergel, D. N.  \& Hernquist, L. 1995, \apj, 451, 598

\bibitem[Paper I]{paper1}
Majewski, S.R.,  Siegel, M.H.,
Kunkel, W.E.,
Reid, I.N.,
Thompson, I.B., 
Landolt, A.U. \&
Johnston, K.V., 1999 \aj, submitted (Paper I)

\bibitem[M98]{m98}
Mateo, M., Olszewski, E. \& Morrison, H., \apjl, in press (M98)

\bibitem[M96]{m96} 
Mateo, M., Mirabal, N., Udalski, A., Szyman\'ski, M., 
Ka{\l}u\`zny, J., Kubiak, M., Krzemi\'nski, W. \& Stanek, 
K. Z. 1996, \apjlett, 458, L13 (M96)

\bibitem[Murali \& Dubinski (1999)]{md99}
Murali, C. \& Dubinski, J. 1999, \aj, in press

\bibitem[Olling \& Merrifield 1998]{om98}
Olling, R. P. \& Merrifield, M. R., 1998, ASP Conf. Ser. 136: Galactic Halos,
p. 219 (Ed. D. Zaritsky)

\bibitem[Vel\'asquez \& White (1995)]{vw95} 
Vel\'asquez, H. \& White, S. 1995, \mnras, 275, L23

\bibitem[Zhao (1998)]{z98}
Zhao, H.S. 1998, \apjl, 500, L149

\end{thebibliography}
\end{document}